\documentclass{appolb}
\usepackage{epsfig}

\def\bge{\begin{equation}}
\def\ene{\end{equation}}
\def\bg{\begin{eqnarray}}
\def\en{\end{eqnarray}}

\def\ubar{{\bar{u}}}
\def\dbar{{\bar{d}}}
\def\sbar{{\bar{s}}}

\def\bge{\begin{equation}}
\def\ene{\end{equation}}
\def\bg{\begin{eqnarray}}
\def\en{\end{eqnarray}}

\def\ubar{{\bar{u}}}
\def\dbar{{\bar{d}}}
\def\sbar{{\bar{s}}}

\begin{document}
\title{Anomalous glue, $\eta$ and $\eta'$ mesons
\thanks{Presented at the Symposium on Meson Physics, Cracow, October 1-4 2008.}
}
\author{Steven D. Bass
\address{Institute for Theoretical Physics, University of Innsbruck, \\
Technikerstrasse 25, A6020 Innsbruck, Austria}
}
\maketitle
\begin{abstract}
Axial U(1) dynamics are characterised by large OZI violations.
Here we review the phenomenology of $\eta$ and $\eta'$
production and decay processes, 
and its connection to the anomalous glue 
that generates a large part of the masses of these pseudoscalar mesons.
\end{abstract}

\section{Introduction}

The flavour-singlet $J^P = 1^+$ channel
is characterised by large OZI violation:
the masses of the $\eta$ and $\eta'$ mesons are much greater 
than the values they would have if these mesons were 
pure Goldstone bosons associated with spontaneously broken chiral 
symmetry \cite{uppsala}.
This extra mass is induced by non-perturbative gluon dynamics and 
the QCD axial anomaly \cite{zuoz}.
How is this anomalous glue manifest in $\eta$ and $\eta'$ 
production and decay processes and in their interactions with nuclear matter ?
These processes are being studied in experiments from threshold
\cite{moskal}
through to high-energy collisions
where 
anomalously large branching ratios have been
observed for $D_s$ and $B$-meson decays to an $\eta'$ plus additional 
hadrons \cite{frere,hf}.
The QCD axial anomaly is also important in discussion of the proton 
spin puzzle \cite{spin}.

Here we outline the key issues.

\section{QCD considerations}

Spontaneous chiral symmetry breaking in QCD is associated with a 
non-vanishing chiral condensate
\begin{equation}
\langle \ {\rm vac} \ | \ {\bar q} q \ | \ {\rm vac} \ \rangle < 0
.
\label{eq5}
\end{equation}
The non-vanishing chiral condensate also spontaneously breaks the axial
U(1) symmetry so, naively, in the two-flavour theory one expects an
isosinglet pseudoscalar degenerate with the pion.
The lightest mass isosinglet is the $\eta$ meson, which has a mass of
547.75 MeV.

The puzzle deepens when one considers SU(3).
Spontaneous chiral symmetry breaking suggests an octet of 
would-be Goldstone bosons:
the octet associated with chiral $SU(3)_L \otimes SU(3)_R$
plus a singlet boson associated with axial U(1)
--- each with mass squared $m^2_{\rm Goldstone} \sim m_q$.
The physical $\eta$ and $\eta'$ masses
are
about 300-400 MeV too big to fit in this picture.
One needs extra mass in the singlet channel
associated with
non-perturbative topological gluon configurations and
the QCD axial anomaly.
The strange quark mass induces considerable $\eta$-$\eta'$ mixing.
For free mesons
the $\eta - \eta'$ mass matrix (at leading order in the chiral
expansion) is
\begin{equation}
M^2 =
\left(\begin{array}{cc}
{4 \over 3} m_{\rm K}^2 - {1 \over 3} m_{\pi}^2  &
- {2 \over 3} \sqrt{2} (m_{\rm K}^2 - m_{\pi}^2) \\
\\
- {2 \over 3} \sqrt{2} (m_{\rm K}^2 - m_{\pi}^2) &
[ {2 \over 3} m_{\rm K}^2 + {1 \over 3} m_{\pi}^2 + {\tilde m}^2_{\eta_0} ]
\end{array}\right)
.
\label{eq10}
\end{equation}
Here ${\tilde m}^2_{\eta_0}$ is the gluonic mass term which has a
rigorous interpretation through the Witten-Veneziano mass formula
\cite{witten,vecca}
and which
is associated with non-perturbative gluon
topology, related perhaps to confinement \cite{ks} or instantons
\cite{thooft}.
The masses of the physical $\eta$ and $\eta'$ mesons are found
by diagonalizing this matrix, {\it viz.}
\begin{eqnarray}
| \eta \rangle &=&
\cos \theta \ | \eta_8 \rangle - \sin \theta \ | \eta_0 \rangle
\\ \nonumber
| \eta' \rangle &=&
\sin \theta \ | \eta_8 \rangle + \cos \theta \ | \eta_0 \rangle
\label{eq11}
\end{eqnarray}
where
\begin{equation}
\eta_0 = \frac{1}{\sqrt{3}}\; (u\ubar + d\dbar + s\sbar),\quad
\eta_8 = \frac{1}{\sqrt{6}}\; (u\ubar + d\dbar - 2 s\sbar) 
.
\label{mixing2}
\end{equation}
One obtains values for the $\eta$ and $\eta'$ masses:
\begin{eqnarray}
m^2_{\eta', \eta} 
& &= (m_{\rm K}^2 + {\tilde m}_{\eta_0}^2 /2)
\nonumber \\
& & \pm {1 \over 2}
\sqrt{(2 m_{\rm K}^2 - 2 m_{\pi}^2 - {1 \over 3} {\tilde m}_{\eta_0}^2)^2
   + {8 \over 9} {\tilde m}_{\eta_0}^4} 
.
\nonumber \\
\label{eq12}
\end{eqnarray}
The physical mass of the $\eta$ and the octet mass
$
m_{\eta_8} = \sqrt{ {4 \over 3} m_{\rm K}^2 - {1 \over 3} m_{\pi}^2 }
$
are numerically close, within a few percent.
However, to build a theory of the $\eta$ on the octet
approximation
risks losing essential physics associated with the singlet component.
Turning off the gluonic term, one finds the expressions
$m_{\eta'} \sim \sqrt{2 m_{\rm K}^2 - m_{\pi}^2}$
and
$m_{\eta} \sim m_{\pi}$.
That is, without extra input from glue, in the OZI limit,
the $\eta$ would be approximately an isosinglet light-quark state
(${1 \over \sqrt{2}} | {\bar u} u + {\bar d} d \rangle$)
degenerate with the pion and
the $\eta'$ would be a strange-quark state $| {\bar s} s \rangle$
--- mirroring the isoscalar vector $\omega$ and $\phi$ mesons.

Taking the value ${\tilde m}_{\eta_0}^2 = 0.73$GeV$^2$ in the 
leading-order 
mass formula, Eq.(\ref{eq12}), 
gives agreement with the physical masses at the 10\% level.
This value is obtained by summing over the two eigenvalues 
in Eq.(5):
$
m_{\eta}^2 + m_{\eta'}^2 = 2 m_K^2 + {\tilde m}_{\eta_0}^2 
$
and substituting 
the physical values of $m_{\eta}$, $m_{\eta'}$ and $m_K$ \cite{vecca}.
The 
corresponding
$\eta - \eta'$
mixing angle $\theta \simeq - 18^\circ$ 
is within the range $-17^\circ$ to $-20^\circ$ obtained
from a study of various decay processes in \cite{gilman,frere}.
\footnote{
Closer agreement with the physical masses can be obtained 
by introducing
the singlet decay constant
$F_0 \neq F_{\pi}$ and including higher-order mass terms in the chiral
expansion
\cite{leutwyler,feldmann}.
}
The key point of Eq.(5) is that mixing and gluon dynamics play a crucial 
role 
in both the $\eta$ and $\eta'$ masses.

\section{The axial anomaly and ${\tilde m}_{\eta_0}^2$}

The flavour-singlet part of $\eta$ and $\eta'$ mesons couples 
to the flavour-singlet axial-vector current $J_{\mu 5}$
\begin{equation}
J_{\mu 5} =
\biggl( \bar{u}\gamma_\mu\gamma_5u
                  + \bar{d}\gamma_\mu\gamma_5d
                  + \bar{s}\gamma_\mu\gamma_5s \biggr) .
\end{equation} 
In classical field theory this current would 
be the partially conserved
Noether current associated with axial U(1) symmetry.
In QCD
renormalization effects mean that 
$J_{\mu 5}$ satisfies the anomalous divergence equation 
\begin{equation}
\partial^\mu J_{\mu5} = 
6 \partial^\mu K_\mu 
+
\sum_{k=1}^{3} 2 i m_k \bar{q}_k \gamma_5 q_k 
\label{eqf102}
\end{equation}
where
\begin{equation}
K_{\mu} = {g^2 \over 32 \pi^2}
\epsilon_{\mu \nu \rho \sigma}
\biggl[ A^{\nu}_a \biggl( \partial^{\rho} A^{\sigma}_a
- {1 \over 3} g
f_{abc} A^{\rho}_b A^{\sigma}_c \biggr) \biggr]
\label{eqf103}
\end{equation}
is the gluonic Chern-Simons current.
Here $A^{\mu}_a$ is the gluon field and
\begin{equation}
Q = \partial^{\mu} K_{\mu}
= {g^2 \over 32 \pi^2} G_{\mu \nu} {\tilde G}^{\mu \nu}
\end{equation}
is the (gauge-invariant) topological charge density,
$G_{\mu \nu}$ is the gluon field tensor and
${\tilde G}^{\mu \nu} = {1 \over 2} \epsilon^{\mu \nu \alpha \beta}
G_{\alpha \beta}$.
Its integral over space
$\int \ d^4 z \ Q = n$ 
measures the gluonic winding number \cite{crewther},
which is an integer for (anti-)instantons and which 
vanishes in perturbative QCD.
Eq.(\ref{eqf102}) allows us to define a partially conserved current
$
J_{\mu 5} = J_{\mu 5}^{\rm con} + 2f K_{\mu}
$,
{\it viz.}
$
\partial^\mu J^{\rm con}_{\mu5}
= \sum_{i=1}^{3} 2im_i \bar{q}_i\gamma_5 q_i
$.
When we make a gauge transformation $U$
the gluon field transforms as
$
A_{\mu} \rightarrow U A_{\mu} U^{-1} + {i \over g} (\partial_{\mu} U) U^{-1}
$
and the operator $K_{\mu}$
transforms as
\begin{eqnarray}
K_{\mu} \rightarrow K_{\mu}
&+&
 i {g \over 8 \pi^2} \epsilon_{\mu \nu \alpha \beta}
\partial^{\nu}
\biggl( U^{\dagger} \partial^{\alpha} U A^{\beta} \biggr)
\nonumber \\
&+& 
+
{1 \over 24 \pi^2} \epsilon_{\mu \nu \alpha \beta}
\biggl[
(U^{\dagger} \partial^{\nu} U)
(U^{\dagger} \partial^{\alpha} U)
(U^{\dagger} \partial^{\beta} U)
\biggr]
.
\label{eqf106}
\end{eqnarray}
In general, matrix elements of $K_{\mu}$ are gauge dependent
This means that one has to be careful
writing matrix elements of $J_{\mu 5}$ 
as the sum of
(measurable)
``quark'' and ``gluonic'' contributions.

\subsection{The U(1) effective Lagrangian for low-energy QCD}

Independent of the detailed QCD dynamics one can construct low-energy 
effective chiral Lagrangians which include the effect of the anomaly 
and axial U(1) symmetry, 
and use these Lagrangians to study low-energy processes involving the 
$\eta$ and $\eta'$.

The physics of axial U(1) degrees of freedom is described 
by the 
U(1)-extended low-energy effective Lagrangian \cite{vecca}.
In its simplest form this reads
\begin{eqnarray}
{\cal L} =
{F_{\pi}^2 \over 4}
{\rm Tr}(\partial^{\mu}U \partial_{\mu}U^{\dagger})
+
{F_{\pi}^2 \over 4} {\rm Tr} M \biggl( U + U^{\dagger} \biggr)
\nonumber \\
+ {1 \over 2} i Q {\rm Tr} \biggl[ \log U - \log U^{\dagger} \biggr]
+ {3 \over {\tilde m}_{\eta_0}^2 F_{0}^2} Q^2
.
\label{eq20}
\end{eqnarray}
Here 
$
U = \exp \ i \biggl(  \phi / F_{\pi}
                  + \sqrt{2 \over 3} \eta_0 / F_0 \biggr) 
$
is the unitary meson matrix
where
$\phi = \ \sum \pi_a \lambda_a$ 
denotes the octet of would-be Goldstone bosons associated 
with spontaneous chiral $SU(3)_L \otimes SU(3)_R$ breaking
and
$\eta_0$
is the singlet boson.
In Eq.(11) $Q$ denotes the topological charge density;
$M = {\rm diag} [ m_{\pi}^2, m_{\pi}^2, 2 m_K^2 - m_{\pi}^2 ]$
is the quark-mass induced meson mass matrix.
The pion decay constant $F_{\pi} = 92.4$MeV and 
$F_0$ is
the flavour-singlet decay constant,
$F_0 \sim F_{\pi} \sim 100$ MeV \cite{gilman}.

The flavour-singlet potential involving $Q$ is introduced to generate 
the gluonic contribution to the $\eta$ and $\eta'$ masses and
to reproduce the anomaly in the divergence of
the gauge-invariantly renormalised flavour-singlet axial-vector
current.
The gluonic term $Q$ is treated as a background field with no kinetic 
term. It may be eliminated through its equation of motion to generate 
a gluonic mass term for the singlet boson,
{\it viz.}
\begin{equation}
{1 \over 2} i Q {\rm Tr} \biggl[ \log U - \log U^{\dagger} \biggr]
+ {3 \over {\tilde m}_{\eta_0}^2 F_{0}^2} Q^2
\
\mapsto \
- {1 \over 2} {\tilde m}_{\eta_0}^2 \eta_0^2
.
\label{eq23}
\end{equation}
The most general low-energy effective Lagrangian involves a $U_A(1)$
invariant polynomial in $Q^2$.  Higher-order terms in $Q^2$ become
important when we consider scattering processes involving more than
one $\eta'$ \cite{veccb}.
In general, couplings involving $Q$ give OZI violation in physical
observables.

\section{Light-cone wavefunctions and fragmentation functions}

In general, there are gluonic effects in $\eta$ and $\eta'$ 
phenomenology associated with the gluonic potential 
involving the topological charge density in 
the U(1)-extended effective chiral Lagrangian for low energy 
QCD, OZI violation in the intermediate states of reactions involving
flavour-singlet hadrons, and gluonic Fock components
in the $\eta$ and 
$\eta'$ light-cone wavefunctions.
At a theoretical level, 
technical issues include 
separating
leading contributions associated with matrix elements of the singlet
axial vector current 
${\bar \psi} \gamma_{\mu} \gamma_5 \psi$
and higher twist effects associated with $J^P=1^+$
gauge invariant gluonic operators 
like 
$G_{\alpha \beta} i D_{\mu} {\tilde G}^{\alpha \beta}$
in the definition of the $\eta'$ (light-cone) wavefunction
\footnote{There is no gauge-invariant twist-2, spin-one gluonic
operator with $J^P=1^+$.}.
In the first case gluonic effects enter through the topological 
charge density in the anomalous divergence of the singlet current and 
in matrix elements involving the gauge-dependent anomalous 
Chern-Simons current $K_{\mu}$
making any quark-gluon separation subtle and, 
where meaningful, should be defined with respect to a certain choice of 
gauge.
(In perturbation theory and in the light-cone gauge the forward matrix 
 elements of $K_{\mu}$ are invariant under residual gauge degrees of 
 freedom, allowing one to connect these matrix elements with polarised 
 glue in the QCD parton model \cite{spin,etar}. 
 The matrix elements of $K_{\mu}$ are gauge dependent 
 as soon as one moves away from the forward direction.)

Consider the (leading twist) light-cone wavefunctions of the $\eta$ and 
$\eta'$ mesons \cite{feldmann,sjb}.
For the meson $P$ ($\eta$ or $\eta'$),
let
$\Psi_P^i (x, {\vec k}_t)$ 
denote the amplitude for finding a quark-antiquark pair carrying
light-cone momentum fraction $x$ and $(1-x)$ 
and transverse momentum ${\vec k}_t$;
$i$ denotes 
the SU(3) octet or singlet ($i=8$ or 0) component of the wavefunction.
These amplitudes are normalised 
via
\begin{equation}
\int {d^2 {\vec k}_{\perp} \over 16 \pi^3} \int_0^1 dx 
\Psi_P^i (x, {\vec k}_{\perp}) = {f_P^i \over 2 \sqrt{6}} 
\end{equation}
where
\begin{equation}
\langle {\rm vac} | J_{\mu 5}^i | P(p) \rangle = i f_P^i \ p_{\mu}
\end{equation}
with $f_P^i$ the corresponding decay constants \cite{feldmann,sjb}.
Gauge dependence issues arise immediately that one tries 
to separate
a ``$K_{\mu}$ contribution'' 
from matrix elements of the singlet current $J_{\mu 5}$.
If we calculate the hard (perturbative QCD) part of an $\eta$ or $\eta'$
production or decay process
using a gauge-invariant 
scheme like $\overline{\rm MS}$,
then the anomalous glue associated with 
the QCD axial anomaly 
will be
included in the ``quark-antiquark part'' of the $\eta$ or $\eta'$ 
wavefunction with 
the quark-antiquark pair feeling the effect of 
the OZI violating gluonic potential associated with $Q$
and (possible) strong coupling 
to glue in the intermediate state of the reaction.
The $\eta$-$\eta'$ mixing angle is built into the light-cone wavefunction.
Separate to glue associated with the QCD axial anomaly, 
one might also consider mixing with the lightest mass $0^-$ glueball.
Possible candidates for this state include the $\eta(1405)$ 
and a glueball predicted by lattice QCD with mass above 2 GeV \cite{pdg}.
Studies of possible gluonic components 
in 
the meson wavefunctions have been carried out in fits to data on
exclusive
$\eta$ and
$\eta'$ production and decay processes \cite{feldmann,etafits}.

Semi-inclusive $\eta$ production in high-energy collisions has been a
topical issue since the pioneering work of Field and Feynman \cite{feynman}.
One finds 
the interesting
result that
the ratio of $\eta$ to $\pi^0$ production 
rises rapidly with the transverse momentum of the produced meson and
levels off at at $R_{\eta/\pi^0} \sim 0.4 - 0.5$ 
above $p_t \sim 3$ GeV
in hadron-hadron collisions
(proton-proton, proton-ion and ion-ion)
independent of the colliding hadron species \cite{rhic},
consistent with the expectations from string fragmentation models.
Studies of $\eta$ and $\eta'$ production in hadron jets at LEP 
were performed 
\cite{lep}.
While the L3 analysis claims to observe an excess of 
$\eta$ production
in gluon jets, neither OPAL nor ALEPH found an excess.
The ratio of $\eta$ to $\pi^0$ multiplicities in quark and gluon 
jets was measured over the range $x= E/E_{\rm beam}$ between 0.1 and 0.5.
Good fits to these ratios 
are
$R_{\eta/\pi^0} = 1.1 x^{0.94}$ 
in quark jets
and
$3.4 x^{1.01} (1-x)$
in gluon jets over the measured region.
$\eta'$ production was observed to be anomalously suppressed 
compared to the expectations of string fragmentation models 
without an additional ``$\eta'$ suppression factor'', 
possibly associated with the mass of the produced $\eta'$.

\section{Low energy $\eta$ and $\eta'$ hadron interactions}

\subsection{Light-mass exotic meson production}

The interactions of the $\eta$ and $\eta'$ with other mesons and 
with nucleons can be studied by coupling the Lagrangian Eq.(11) to 
other particles.
For example,
the OZI violating interaction
$\lambda Q^2 \partial_{\mu} \pi_a \partial^{\mu} \pi_a$
is needed to generate the leading (tree-level)
contribution to the decay $\eta' \rightarrow \eta \pi \pi$
\cite{veccb}.
When iterated in the Bethe-Salpeter equation for meson-meson
rescattering
this interaction yields a dynamically generated exotic state
with quantum numbers $J^{PC} = 1^{-+}$ and mass about 1400 MeV
\cite{bassmarco}.
This suggests a dynamical interpretation of the lightest-mass 
$1^{-+}$ exotic observed at BNL \cite{exoticb} and CERN \cite{exoticc}.

\subsection{Proton-nucleon collisions}

For proton-nucleon collisions one finds a gluon-induced contact 
interaction 
in the 
$pp \rightarrow pp \eta'$ reaction \cite{bass99}:
\begin{equation}
{\cal L}_{\rm contact} =
         - {i \over F_0^2} \ g_{QNN} \ {\tilde m}_{\eta_0}^2 \
           {\cal C} \
           \eta_0 \ 
           \biggl( {\bar p} \gamma_5 p \biggr)  \  \biggl( {\bar p} p \biggr) .
\end{equation}
Here 
$g_{QNN}$ is the 1PI coupling of $Q$ to the nucleon and ${\cal C}$ is a 
second OZI violating coupling.
The physical interpretation of the contact term (15) 
is a ``short distance'' ($\sim 0.2$fm) interaction 
where glue is excited in the interaction region of
the proton-proton collision and 
then evolves to become an $\eta'$ in the final state.
This gluonic contribution to the cross-section 
for $pp \rightarrow pp \eta'$ 
is extra to the contributions associated with 
meson exchange models 
There is no reason, a priori, to expect it to be small.
Since glue is flavour-blind the contact interaction (15) has the same 
size in both 
the $pp \rightarrow pp \eta'$ and $pn \rightarrow pn \eta'$ reactions.
The ratio
$R_{\eta} 
 = \sigma (pn \rightarrow pn \eta ) / \sigma (pp \rightarrow pp \eta )$
has been measured
for quasifree $\eta$ 
production from a deuteron target up to 100 MeV above threshold
\cite{pawelsb}.
One finds that $R_{\eta}$ 
is approximately energy-independent 
$\sim 6.5$ 
over the energy range $20 - 100$ MeV
signifying a strong isovector exchange 
contribution to the $\eta$ production mechanism.
In the extreme scenario that the glue-induced production saturated 
the $\eta'$ 
production cross-section, the ratio
$R_{\eta'} =
 \sigma (pn \rightarrow pn \eta' ) / \sigma (pp \rightarrow pp \eta' )$
would go to one
after we correct for the final state interaction between the two outgoing 
nucleons.
Proton-proton data is available from COSY-11 \cite{moskal}; 
the proton-neutron process has been measured and the data is being analysed 
\cite{joanna}.

\subsection{$\eta$ and $\eta'$ interactions with the nuclear medium}

Measurements of the pion, kaon and eta meson masses and their interactions
in finite nuclei provide new constraints on our understanding of dynamical
symmetry breaking in low energy QCD \cite{kienle}.
For the $\eta$
the in-medium mass $m_{\eta}^*$ 
is sensitive to the flavour-singlet
component in the $\eta$, and hence 
to the non-perturbative glue associated with axial U(1) dynamics.
An important source of the in-medium mass modification comes
from light-quarks
coupling to the scalar $\sigma$ mean-field in the nucleus.
Increasing the flavour-singlet component in the $\eta$
at the expense of the octet component gives more attraction,
more binding and a larger value of the $\eta$-nucleon
scattering length, $a_{\eta N}$.
Since the mass shift is approximately proportional to the $\eta$--nucleon 
scattering length, it follows that that the physical value of $a_{\eta N}$ 
should be larger than if the $\eta$ were a pure octet state.

Meson masses in nuclei are determined from the scalar induced contribution 
to the meson propagator evaluated at zero three-momentum, ${\vec k} =0$, in 
the nuclear medium.
Let $k=(E,{\vec k})$ and $m$ denote the four-momentum and mass of the meson 
in free space.
Then, one solves the equation
\begin{equation}
k^2 - m^2 = {\tt Re} \ \Pi (E, {\vec k}, \rho)
\end{equation}
for ${\vec k}=0$
where $\Pi$ is the in-medium $s$-wave meson self-energy.
Contributions to the in medium mass come from coupling to the scalar 
$\sigma$ field in the nucleus in mean-field approximation,
nucleon-hole and resonance-hole excitations in the medium.
The $s$-wave self-energy can be written as \cite{ericson}
\begin{equation}
\Pi (E, {\vec k}, \rho) \bigg|_{\{{\vec k}=0\}}
=
- 4 \pi \rho \biggl( { b \over 1 + b \langle {1 \over r} \rangle } \biggr) .
\end{equation}
Here $\rho$ is the nuclear density,
$
b = a ( 1 + {m \over M} )
$
where 
$a$ is the meson-nucleon scattering length, $M$ is the nucleon mass and
$\langle {1 \over r} \rangle$ is
the inverse correlation length,
$\langle {1 \over r} \rangle \simeq m_{\pi}$ 
for nuclear matter density \cite{ericson}.
($m_{\pi}$ is the pion mass.) 
Attraction corresponds to positive values of $a$.
The denominator in Eq.(17) is the Ericson-Ericson-Lorentz-Lorenz
double scattering correction.

What should we expect for the $\eta$ and $\eta'$ ?

To investigate what happens to ${\tilde m}^2_{\eta_0}$ in the medium
we first couple
the $\sigma$ 
(correlated two-pion)
mean-field in nuclei
to the topological charge density $Q$ by adding the Lagrangian term
\begin{equation}
{\cal L}_{\sigma Q} =
Q^2 \ g_{\sigma}^Q \sigma
\label{eq27}
\end{equation}
where 
$g_{\sigma}^Q$ denotes coupling to the $\sigma$ mean field
--
that is, we
consider an in-medium renormalization of the coefficient of $Q^2$
in the effective chiral Lagrangian \cite{bt05}.
We can eliminate
$Q$ through its equation of motion (following Eq.(12)).
The gluonic mass term for the singlet boson then becomes
\begin{equation}
{\tilde m}^2_{\eta_0}
\mapsto
{\tilde m}^{*2}_{\eta_0}
=
{\tilde m}^2_{\eta_0}
\ { 1 + 2 x \over (1 + x)^2 }
\ < {\tilde m}^2_{\eta_0}
\label{eq28}
\end{equation}
where
\begin{equation}
x =
{1 \over 3} g_{\sigma}^Q \sigma \ {\tilde m}^2_{\eta_0} F_0^2.
\label{eq29}
\end{equation}
That is, {\it the gluonic mass term decreases in-medium}
independent of the sign of $g_{\sigma}^Q$ and the medium acts
to partially neutralise axial U(1) symmetry breaking by gluonic effects.

The above discussion is intended to motivate the {\it existence} of 
medium modifications to ${\tilde m}^2_{\eta_0}$ in QCD.
\footnote{
In the chiral limit the singlet
analogy to the Weinberg-Tomozawa
term does not vanish because of the anomalous glue terms.
Starting from the simple Born term one finds
anomalous gluonic contributions
to the singlet-meson nucleon scattering length
proportional to ${\tilde m}^2_{\eta_0}$ and ${\tilde m}_{\eta_0}^4$
\cite{bassww}.
}
However, a rigorous calculation of $m_{\eta}^{*}$ from QCD 
is beyond present theoretical technology. 
Hence, one has to look to QCD motivated models and phenomenology for
guidance about the numerical size of the effect.
The physics described in 
Eqs.(2-5) tells us that the simple octet approximation may not suffice.

This physics has been investigated by Bass and Thomas \cite{bt05}.
Phenomenology is used
to estimate the size of the effect in the $\eta$
using
the Quark Meson Coupling model (QMC) of hadron properties in the nuclear 
medium \cite{etaqmc}.
Here one uses the large $\eta$ mass 
(which in QCD is induced by mixing and the gluonic mass term)
to motivate taking an MIT Bag 
description
for the $\eta$ wavefunction, and
then coupling the light (up and down)
quark and antiquark fields in the $\eta$ to the scalar $\sigma$
field
in the nucleus working in mean-field approximation \cite{etaqmc}.
The strange-quark component of the wavefunction does not couple
to the $\sigma$ field and $\eta-\eta'$ mixing is readily built into the model.

Increasing the mixing angle increases the amount of singlet 
relative to octet components in the $\eta$.
This produces greater attraction through increasing 
the amount of light-quark compared to strange-quark 
components in the $\eta$
and a reduced effective mass.
Through Eq.(17), increasing the mixing angle 
also increases 
the
$\eta$-nucleon scattering length $a_{\eta N}$.
The model results are shown in Table 1.
The key observation is that $\eta - \eta'$ mixing leads to a factor of 
two increase in the mass-shift and in the scattering length obtained in 
the model.
This result may explain why values of $a_{\eta N}$ extracted from 
phenomenological fits to experimental data where the $\eta-\eta'$ 
mixing angle is unconstrained 
give larger values than those predicted 
in theoretical models where the $\eta$ is treated as a pure octet state.

\begin{table}[t!]
\begin{center}
\caption{
Physical masses fitted in free space, 
the bag 
masses in medium at normal nuclear-matter 
density,
$\rho_0 = 0.15$ fm$^{-3}$, 
and corresponding meson-nucleon scattering lengths
(calculated at the mean-field level 
 with the Ericson-Ericson-Lorentz-Lorenz factor switched off).
}
\label{bagparam}
\begin{tabular}[t]{c|lll}
\hline
&$m$ (MeV) 
& $m^*$ (MeV) & ${\tt Re} a$ (fm)
\\
\hline
$\eta_8$  &547.75  
& 500.0 &  0.43 \\
$\eta$ (-10$^o$)& 547.75  
& 474.7 & 0.64 \\
$\eta$ (-20$^o$)& 547.75  
& 449.3 & 0.85 \\
$\eta_0$  &      958 
& 878.6  & 0.99 \\
$\eta'$ (-10$^o$)&958 
& 899.2 & 0.74 \\
$\eta'$ (-20$^o$)&958 
& 921.3 & 0.47 \\
\hline
\end{tabular}
\end{center}
\end{table}

The density dependence of the mass-shifts in the QMC model is discussed
in Ref.\cite{etaqmc}.
Neglecting the Ericson-Ericson term, the mass-shift is approximately
linear
For densities $\rho$ between 0.5 and 1 times $\rho_0$ (nuclear
matter density) we find
\begin{equation}
m^*_{\eta} / m_{\eta} \simeq 1 - 0.17 \rho / \rho_0
\end{equation}
for the mixing angle $-20^\circ$.
The scattering lengths extracted from this analysis are density independent 
to within a few percent over the same range of densities.

\vspace{1.0cm}

{\bf Acknowledgements} \\

I thank C. Aidala and P. Kroll for helpful discussions
and P. Moskal
for the invitation to talk at this stimulating meeting. 
The work of SDB is supported by the Austrian Research Fund, FWF, through 
contract P20436.

\end{document}